\documentclass[%
 aip,
 amsmath,amssymb,
preprint,%
]{revtex4-1}

\usepackage{graphicx}
\usepackage{dcolumn}
\usepackage{bm}

\usepackage[utf8]{inputenc}
\usepackage[T1]{fontenc}
\usepackage{mathptmx}
\usepackage{etoolbox}

\makeatletter
\def\@email#1#2{%
 \endgroup
 \patchcmd{\titleblock@produce}
  {\frontmatter@RRAPformat}
  {\frontmatter@RRAPformat{\produce@RRAP{*#1\href{mailto:#2}{#2}}}\frontmatter@RRAPformat}
  {}{}
}%
\makeatother
\begin{document}


\title{The photodissociation dynamics and ultrafast electron diffraction image of cyclobutanone from the surface hopping dynamics simulation}
\author{Jiawei Peng}
\affiliation{School of Chemistry, South China Normal University, Guangzhou 510006, China}
\affiliation{MOE Key Laboratory of Environmental Theoretical Chemistry, South China Normal University, Guangzhou 510006, China}
\affiliation{SCNU Environmental Research Institute, Guangdong Provincial Key Laboratory of Chemical Pollution and Environmental Safety, School of Environment, South China Normal University, Guangzhou 510006, China}

\author{Hong Liu}
\email{hongliu@m.scnu.edu.cn.}
\affiliation{MOE Key Laboratory of Environmental Theoretical Chemistry, South China Normal University, Guangzhou 510006, China}
\affiliation{SCNU Environmental Research Institute, Guangdong Provincial Key Laboratory of Chemical Pollution and Environmental Safety, School of Environment, South China Normal University, Guangzhou 510006, China}

\author{Zhenggang Lan}
\email{zhenggang.lan@m.scnu.edu.cn; zhenggang.lan@gmail.com.}
\affiliation{MOE Key Laboratory of Environmental Theoretical Chemistry, South China Normal University, Guangzhou 510006, China}
\affiliation{SCNU Environmental Research Institute, Guangdong Provincial Key Laboratory of Chemical Pollution and Environmental Safety, School of Environment, South China Normal University, Guangzhou 510006, China}

\date{\today}

\begin{abstract}
The comprehension of nonadiabatic dynamics in polyatomic systems relies heavily on the simultaneous advancements in theoretical and experimental domains. The gas-phase electron diffraction (GUED) technique has attracted widespread attention as a promising tool for observing the photochemical and photophysical features at all-atomic level with high temporal and spatial resolutions. In this work, the GUED spectra were predicted to perform a double-blind test of accuracy in excited-state simulation for cyclobutanone based on the trajectory surface hopping method, with respect to the benchmark data obtained by upcoming MeV-GUED experiments at the Stanfold Linear Accelerator Laboratory. The results show that the ultrafast nonadiabatic dynamics occurs in the photoinduced dynamics, and two $C2$ and $C3$ channels play dominant roles in the nonadiabatic reactions of cyclobutanone. The simulated UED signal can be directly interpreted by atomic movements, providing a unique view to monitor the time-dependent evolution of the molecular structure in the femtosecond dynamics.
\end{abstract}

\maketitle


\section{INTRODUCTION}
Photoinduced nonadiabatic processes play an important role in numerous photochemical, photophysical, and photobiological reactions.\cite{cheng2009dynamics,domcke2004conical,domcke2011conical,gozem2017theory,curchod2018ab,crespo2018recent} The understanding of nonadiabatic dynamics of polyatomic systems is strongly dependent on the common progress in both theoretical and experimental fields.\cite{pollard1992analysis,zewail2000femtochemistry,stolow2004femtosecond,domcke2011conical,kowalewski2017simulating,maiuri2019ultrafast,conti2020ultrafast,polley2021two,yan2021theoretical} 
Owning to the development of the ultrasfast laser facilities and novel detection devices, experimentalists developed various time-resolved spectroscopic techniques to observe and track real-time microscopic movement in molecules,\cite{mukamel1995principles,domcke2011conical,pollard1992analysis,zewail2000femtochemistry,stolow2004femtosecond} while it is not trivial to discovery the corresponding relationship between the intrinsic spectral signals and complex molecular motions.\cite{mukamel1995principles} 
Therefore, theoretical simulation of nonadiabatic dynamics becomes an indispensable tool that complements experimental techniques to understand photochemical and photophysical phenomena.\cite{ben1998ab,domcke2004conical,domcke2011conical,crespo2018recent,gozem2017theory,curchod2018ab,maiuri2019ultrafast,conti2020ultrafast}
In recent decades, many academic advances have been continuously made in this field. Different theoretical methods have been developed, which allow us to directly simulate the excited-state
processes.\cite{crespo2018recent,curchod2018ab,gonzalez2020quantum} 
These theoretical efforts try to explain the experimental observations and greatly deepen the understanding of photochemical and photophysical reactions.\cite{domcke2004conical,domcke2011conical}
On this basis, the employment of various dynamics methods gives us the possibility to predict unknown complex photoinduced reactions and to grasp physical insights behind the new experimental observations, which would pave the way to rationally design the light-driven molecular systems and reform the implications in many fields, including bioimaging, optogenetics, renewable solar energy, and photochemical syntheses.

Taking a simple cyclic ketone with the four-membered ring as a typical example, cyclobutanone ($C_4H_6O$) received the great research interests from both  experimental and theoretical point of views,
\cite{denschlag1968benzene,lee1968photoactivation,scharpen1968microwave,lee1969tracer,whitlock1971electronic,hemminger1971predissociation,lee1971unusual,lee1971fluorescence,hemminger1972fluorescence,hemminger1973laser,izawa1973recoil,breuer1975unimolecular,stigliani1976structure,tang1976laser,harrison1980absorption,causley1980electric,john1983nonequilibrated,tamagawa1983molecular,baba1984s,scheer1984study,nicol1986pulsed,guckert1986application,trentelman1990193,alonso1992centimeter,zhang1994jet,zhao1996classical,ruiz1996simplex,smeyers1997half,brown1998optically,tang1998intramolecular,munrow1999determination,diau2001femtochemistry,carter2001new,melandri2005water,johnson2008cyclobutanone,pirrung2009multicomponent,kuhlman2012symmetry,kuhlman2012coherent,zou2012effect,centore2012series,secci2013stereocontrolled,xia2015excited,liu2016new,kao2020effects}
as it is not only the key initial source for total syntheses of many industrial compounds, but also a basic component for the manufacture of bioactive molecules and medicines.\cite{mulzer2005natural,fleck2008total}
Moreover, the cyclobutanone itself plays as a prototype to test different experimental and theoretical methods, due to its structure simplicity.  

%

Experimentally, the emerging spectroscopic techniques were employed to observe the photochemical and photophysical features of cyclobutanone.\cite{scharpen1968microwave, whitlock1971electronic, hemminger1971predissociation, hemminger1972fluorescence, stigliani1976structure, causley1980electric, tamagawa1983molecular, alonso1992centimeter, pollard1992analysis, zhang1994jet, diau2001femtochemistry, melandri2005water,kao2020effects} Zewail and co-workers studied the absorption spectra of cyclobutanone,\cite{diau2001femtochemistry} in which two electronic absorption bands were found in the experiment. One covers the region from 330 to 240 $nm$, characterizing the 
$n\rightarrow\pi^*$ excitation from the ground state ($S_0$) to the lowest excited state ($S_1$).
The other band appears in the range of 206 to 182 $nm$, which involves the the $n\rightarrow 3s$ Rydberg transition from $S_0$ to the second singlet excited state ($S_2$).\cite{whitlock1971electronic}
Furthermore, Zewail and co-workers also applied the femtosecond time-resolved mass spectrometry to detect the $S_1$ dynamics of cyclobutanone.\cite{diau2001femtochemistry} Recently, Orr-Ewing and co-workers introduced the transient absorption spectroscopy to explore the $S_1$ dynamics of cyclobutanone in solution phase.\cite{kao2020effects} 
For the nonadiabatic process with higher excited states, 
M{\o}ller
and co-workers used the time-resolved mass spectrum and time-resolved photoelectron spectrometry to study the internal conversion from the $S_2$ to $S_1$ states, in which only specific vibrational modes were found to play important roles in the internal conversion of cyclobutanone.\cite{kuhlman2012coherent,kuhlman2013pulling}

Theoretically, several efforts have been made to explore the excited-state processes
of cyclobutanone over the last few decades.\cite{ruiz1996simplex,smeyers1997half,diau2001femtochemistry,xia2015excited,liu2016new} 
Zewail and co-workers inferred that the internal conversion from the $S_1$ to $S_0$ states is an ultrafast process within 50 fs, while the intersystem crossing between the $S_1$ and the lowest triplet excited state ($T_1$) states occurs on a longer time scale.\cite{diau2001femtochemistry}
Cui and co-workers studied the ring-opening mechanism and excited-state deactivation channels of cyclic ketones,\cite{xia2015excited} in which the $C(\alpha)-C$ cleavage of cyclobutanone was discovered in the $S_1$ state once a modest energy barrier is overcome.
To explore its nonadiabatic dynamics, the internal conversion process from the $S_2$ to $S_1$ states was once investigated by 
M{\o}ller
and coworkers using the multiconfiguration time-dependent Hartree method based on the five-dimensional vibronic coupling Hamiltonian,
where the motions of the out-of-plane carbonyl and ring puckering were suggested to have dominant influences on the coupling between the $S_1$ and $S_2$ states.\cite{kuhlman2012coherent} 
The dynamic behaviors of cyclobutanone starting from $S_1$ state were investigated by Fang and co-workers using the \textit{ab initio} multiple spawning dynamics method,\cite{liu2016new} 
which also confirms that the $S_1$ dynamics is an ultrafast process. The reaction channel can be controlled by different excitation wavelengths, where the nonergodic behavior in the $S_1$ dynamics plays an important role in the nonadiabatic reaction.

Until now, the basic understandings of the photoinduced dynamics 
for cyclobutanone were established as follows. 
First, the internal conversion of the $S_2$ to $S_1$ states is a very efficient process. On the one hand, the conical-intersection regions are more accessible due to the higher energies when the $S_2$ state excited. On the other hand, the ring-puckering mode with the low frequency 
plays a very important role in the $S_2 \rightarrow S_1$ nonadiabatic transition.
Two types of products involved in different pathways would be generated after the internal conversion to the $S_0$ 
state, where the $C2$ channel includes ethylene ($C_2H_4$) and ethenone ($CH_2CO$), the $C3$ channel gives propylene ($C_3H_6$) and carbon monoxide ($CO$).\cite{denschlag1968benzene,lee1968photoactivation} The branching ratio between different products depends on the excitation wavelength.\cite{lee1969tracer, tang1976laser}
Although these works have greatly advanced our knowledge of cyclobutanone photodynamics, the comprehensive view of photoinduced dynamics of cyclobutanone at the full atomic level is still not well constructed, when it starts from the $S_2$ state.


Recently, the gas-phase electron diffraction technique (GUED) has attracted widespread attention as a promising tool to determine the gas-phase molecular structure,\cite{centurion2022ultrafast} where the molecular properties are not modified during the probe. This kind of experiments is sensitive to the spatial distribution of nuclei and electrons, and the temporal resolution has been continuously improved in the last ten years. In particular, with the electron gun equipped with megaelectronvolt energy (MeV), the MeV-GUED experiment can be used to explore the femtosecond dynamics by providing spatial resolution of the nuclear motions and observing the reflected changes in the electronic structure. Yang et al. first designed the MeV-GUED experiment to investigate coherent nuclear motion in isolated iodine molecules,\cite{yang2016diffractive} which demonstrates that the MeV-GUED experiment has the capability to accurately capture a vibrational wavepacket at the atomic level, as evidenced by its comparison with the simulated results. Furthermore, subsequent experimental and theoretical investigations have explored the behavior of more complex molecules such as nitrogen, $CF_3I$, $C_2F_4I_2$ and 1,3-$C_6H_8$.\cite{yang2016diffractiven,yang2018imaging,wilkin2019diffractive,wolf2019photochemical} These studies have further demonstrated the effectiveness and strength of this technique. Inspiringly, the MeV-GUED experiment for cyclobutanone is scheduled to be performed at the Stanfold Linear Accelerator Laboratory in the near future, in which a gas sample will be excited with 200 $nm$ light ($\approx$80 fs cross-correlation) at about 1 mbar pressure. The electron diffraction images will then be acquired using a time resolution of 150 $fs$ and a spatial resolution of 0.6 $\overset{\circ}{A}$, the scattering vector $s$ covers a range of 1-10 $\overset{\circ}{A}$$^{-1}$. This not only provides a novel way to investigate the nonadiabatic dynamics of cyclobutanone starting from the $S_2$ state, but also offers an opportunity for a double-blind test of accuracy in excited-state simulations based on various theoretical methods.

In this work, the on-the-fly trajectory surface hopping (TSH) method\cite{tully1990molecular} with the extended multi-state complete active space second order perturbation theory\cite{shiozaki2011communication} (XMS-CASPT2) was performed to study the excited-state nonadiabatic dynamics of cyclobutanone starting from the $S_2$ state, in which the TSH method was chosen due to a good balance of accuracy and computational costs.\cite{crespo2018recent} As speculated previously, the crossing between the system between $S_1$ and the first triplet state occurs at $\sim$5 $ns$,\cite{diau2001femtochemistry} and this time scale is difficult to compete with the much faster direct dissociation channel above the breaking threshold $C(\alpha)-C$ of the $S_1$ state. Therefore, only $S_2$, $S_1$ and $S_0$ were considered in the dynamics simulation. In comparison with the GUED experiment, the diffraction images for different time delays were predicted in terms of the TSH trajectories. The results show that the photodynamics of cyclobutanone occurs on an ultrafast timescale. Most importantly, the simulated GUED signal can be directly interpreted by atomic movements, giving a unique view to monitor the time-dependent evolution of the molecular structure in the nonadiabatic dynamics with rather high temporal and spatial resolutions. This study deepens our understanding of the photochemical process in the cyclobutanone, and can provide inspirations for investigating the photochemical and photophysical processes of other cyclic ketones.

\section{Methods and COMPUTATIONAL DETAILS}
The ground-state geometry optimization and frequency analysis of cyclobutanone were performed at the B3LYP/6-31G* level with the Gaussian 16 package.\cite{frisch2016gaussian} The excited-state equilibrium structures for the $S_1$ ($S_1$-min) and $S_2$ ($S_2$-min) states were determined by the XMS-CASPT2 method with the def2-SVPD basis set (XMS(3)-CASPT2(10, 8)/def2-SVPD), where three states were averaged and 10 electrons in 8 orbitals were selected to define the active space, consisting of $\pi$ and $\pi^*$ orbitals associated with the $C=O$ bond, $\sigma$ and $\sigma^*$ orbitals associated with the $C-C$ bond in the four-membered ring, and non-bonding orbital at the $O$ atom. 
The excited-state properties at the ground-state minima ($S_0$-min) were obtained by the XMS(3)-CASPT2(10, 8)/def2-SVPD and XMS(3)-CASPT2(10, 8)/aug-cc-pVDZ levels with the same active space. The absorption spectrum was simulated employing the nuclear ensemble approach,\cite{crespo2014spectrum} where the geometries were sampled on the basis of the Wigner distribution\cite{wigner1932quantum} of the lowest vibrational state at $S_0$-min. The corresponding vertical excitation energies and oscillator strengths were then calculated at the XMS(3)-CASPT2(10, 8)/def2-SVPD and XMS(3)-CASPT2(10, 8)/aug-cc-pVDZ levels, respectively. Finally, 100 structures were averaged and the Gaussian function with the standard deviation $\delta = 0.1$ $eV$ was used to broaden the spectra.

The ultrafast nonadiabatic dynamics of cyclobutanone was simulated using the on-the-fly TSH method with Tully's fewest-switches algorithm\cite{tully1990molecular} at the XMS(3)-CASPT2(10, 8)/def2-SVPD level. The initial nuclear conditions (geometries and velocities) of different trajectories were generated from the Wigner distribution of the lowest vibrational state at $S_0$-min. In the nonadiabatic dynamics simulation, the velocity-Verlet algorithm was utilized to solve the classical Newton equation of motion, and the time step of nuclear motion was selected as 0.5 $fs$. The unitary propagation approach was employed to solve the time-dependent electronic Schrödinger equation of motion, and the time step of the electronic motion was chosen as 0.005 $fs$. The overcoherence problem was corrected by the method proposed by Granucci et al.,\cite{granucci2007critical} and the parameter was designed as 0.1 $Hartree$ as normally suggested. Once the hopping occurs in the propagation, the velocity rescaling is performed along the direction of the nonadiabatic coupling to ensure energy conservation.\cite{tully1990molecular} In the case of the frustrated hop, the velocity component also needs to be reversed along the nonadiabatic coupling vector.\cite{tully1990molecular} Finally, a total of 100 trajectories were performed for the dynamics starting from the $S_2$ state. The entire calculation was completed with our long-standing developed JADE-NAMD package,\cite{du2015fly} where the PESs, nuclear gradients and nonadiabatic coupling vectors were calculated directly at the XMS-CASPT2 level using the BAGEL package.\cite{shiozaki2018bagel}

The electron diffraction pattern was simulated on the basis of the independent atom model (IAM).\cite{centurion2022ultrafast} The total diffraction signal is separated into two components in this framework, including the atomic scattering part $I_{atom}$ and the molecular scattering part $I_{mol}$, which are satisfied with the following equations\cite{brockway1936electron}:
\begin{align}
  I(s) &= I_{mol}(s) + I_{atom}(s), \\
  I_{atom}(s) &= \sum_{i=1}^N f_i^*(s) f_i(s), \\
  I_{mol}(s) &=  \sum_{i=1}^N\sum_{j=1, j\neq u}^N f_{i}^*(s) f_j(s) 
  \exp(-\frac{1}{2} l_{ij}^2 s^2) \frac{\sin(sr_{ij})}{sr_{ij}},
\end{align}
where $s$ refers to the momentum transfer space, and $N$ represents the number of atoms. $f_i(s)$ is the scattering amplitude of the $i$-th atom, which is calculated using the ELSEPA program in this work.\cite{code} $r_{ij}$ represents the distance between the $i$-th and $j$-th atoms, and $l_{ij}$ denotes the root-mean-square vibrational amplitude of the pair between the $i$-th and $j$-th atoms.

Since $I_{mol}(s)$ formally depends on $s^{-5}$ as indicated in Eq. (3),\cite{centurion2022ultrafast} it decreases rapidly with $s$ increasing, and the signal itself varies significantly over multiple orders. Therefore, a more practical way to represent the scattering signal is in terms of the so-called modified intensity $sM(s)$:
\begin{align}
  sM(s) = s \frac{I_{mol}(s)}{I_{atom}(s)}.
\end{align}
On this basis, the distribution function $P(r)$ in the real space is defined as follows:
\begin{align}
  P(r) = r \int_{0}^{\infty} sM(s)\sin(sr) ds.
\end{align}
Physically, $P(r)$ quantifies the probability of observing an atom pair at any distance $r$. However, only finite $s$ conditions can be measured in the experiment, so $P(r)$ is further approximated as follows:
\begin{align}
  P(r) \approx r \int_{s_{min}}^{s_{max}} sM(s)\sin(sr) e^{-\alpha s^2} ds,
\end{align}
where $s_{min}$ and $s_{max}$ are the lower and upper limits of the detector, $e^{-\alpha s^2}$ is a damping factor to remove the steep sine-transform effects at the high $s$ boundary. Moreover, the GUED experiment also focuses on the change in the diffraction pattern of the target molecule as a result of photoexcitation. Therefore, the time-dependent changes in the diffraction pattern are given by
\begin{align}
  \Delta s M(s, t) &= s \frac{\Delta I_{mol}(s, t)}{I_{atom}(s)} 
  = s \frac{\Delta I (s, t)}{I_{atom}(s)}, \\
  \Delta P(r, t) &\approx r \int_{s_{min}}^{s_{max}} 
  \Delta s M(s, t) \sin (sr) e^{-\alpha s^2} ds.
\end{align}

To match the upcoming MeV-GUED experiment as closely as possible, $s_{min}$ and $s_{max}$ were selected as 0 and 10 $\overset{\circ}{A}$$^{-1}$ in the simulation, and the damping factor $\alpha$ was phenomenologically chosen as 0.05 $\overset{\circ}{A}$$^{-1}$. The value of $sM(s, t)$ was calculated as the average of all dynamics trajectories, and $\Delta P(r, t)$ was simulated by removing the scattering intensity for $S_0$-min to investigate different relaxations. The code with basic functionality is provided by the GUED community.\cite{code}


\section{RESULTS}
\subsection{Excited-state features}
\begin{figure}[htbp]
  \includegraphics[width=16cm,height=6.13cm]{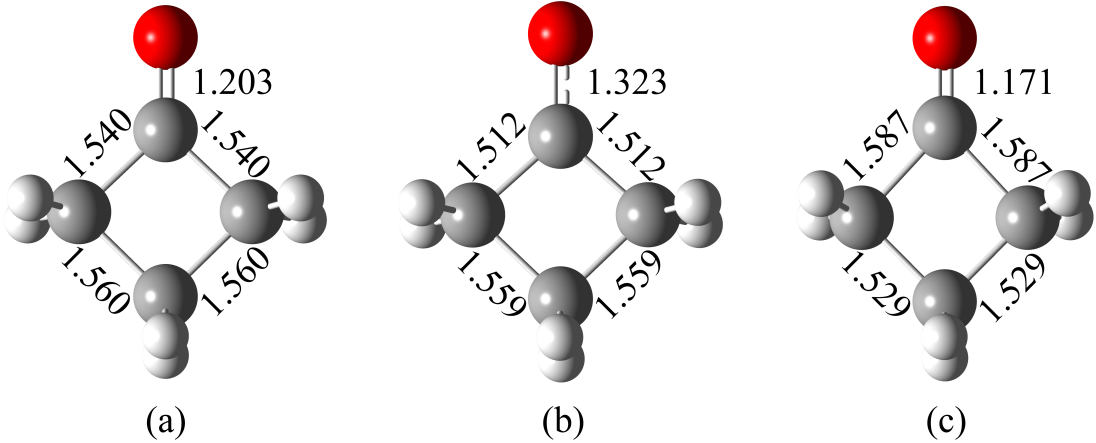}
  \caption{The optimized equlibrium structures of cyclobutanone: 
  (a) $S_0$-min, (b) $S_1$-min ($E^{S_1}_{S_1-min}-E^{S_0}_{S_0-min} = 4.018$ $eV$) and (c) $S_2$-min ($E^{S_2}_{S_2-min}-E^{S_0}_{S_0-min} = 6.511$ $eV$).}
  \label{geom}
\end{figure}

Fig. \ref{geom} shows the equilibrium structures of cyclobutanone in different electronic states with the significant geometric parameters provided. The $S_0$-min exhibits the $C_{2v}$ symmetry, where all $C$ and $O$ atoms are located in the same plane. The lengths of the $C(\alpha)-C$ and $C(\alpha)-C(\beta)$ bonds are measured as 1.540, 1.560 $\overset{\circ}{A}$, respectively, and the $C=O$ distance has 1.203 $\overset{\circ}{A}$ as shown in Fig. \ref{geom}(a). The geometric parameters obtained here are the same as those from previous research.\cite{diau2001femtochemistry} Similar to $S_0$-min, both $S_1$-min and $S_2$-min maintain the planar geometries. However, the $C(\alpha)-C$ bonds of $S_1$-min are reduced by 0.028 $\overset{\circ}{A}$, while the $C=O$ distance is elongated by 0.120 $\overset{\circ}{A}$ in $S_1$-min, as shown in Fig. \ref{geom}(b). The opposite is true for $S_2$-min, in which the lengths of the $C(\alpha)-C$ bonds increase to 1.587 $\overset{\circ}{A}$, while the $C(\beta)-C(\alpha)$ and $C=O$ distances are shortened to 1.529, 1.171 $\overset{\circ}{A}$, respectively, as shown in Fig. \ref{geom}(c).



The excited-state characteristics of cyclobutanone at $S_0$-min are summarized in Table \ref{excited}, in which various basis sets were chosen to perform calculations. The vertical excitation energy of the $S_1$ state at $S_0$-min is 4.290 $eV$, which is within the absorption range of $300$ to $240$ $nm$ as observed experimentally.\cite{hemminger1973laser} This result is consistent with the aug-cc-pVDZ basis set and only has a slight overestimation of 0.083 $eV$. The $S_1$ state calculated here is dominant by the $n \rightarrow \pi^*$ transition, and this state is essentially a dark one, as seen its oscillator strength being zero. The weak absorption peak of this state is fully due to the vibronic intensity borrowing effect. The vertical excitation energy of the $S_2$ state at $S_0$-min is 6.903 $eV$ for the def-SVPD basis set, which is mainly associated with the $n \rightarrow 3s$ Rydberg excitation. And the $S_2$ state is a bright one showing strong oscillator strength. Unlike the def-SVPD basis set, the vertical excitation energy of the $S_2$ state at $S_0$-min is 6.160 $eV$ for the aug-cc-pVDZ basis set.

\begin{table}[htbp]
  \caption{The relative energies and oscillator strengths at the $S_0$ minima, the unit is selected as eV.}
    \begin{ruledtabular}
      \begin{tabular}{cccc}
      State & $S_0$ & $S_1$ &  $S_2$ \\
      \hline
      Energy / def2-SVPD & 0.000   &  4.290      &  6.903     \\
      Oscillator Strength / def2-SVPD &         &  0.000   & 0.622 \\
      Energy / aug-cc-pVDZ & 0.000   &  4.207      &  6.160     \\
      Oscillator Strength / aug-cc-pVDZ &         &  0.000   & 0.736
      \label{excited}
    \end{tabular}
  \end{ruledtabular}
\end{table}



\begin{figure}[htbp]
  \includegraphics[width=12.5cm,height=5.2cm]{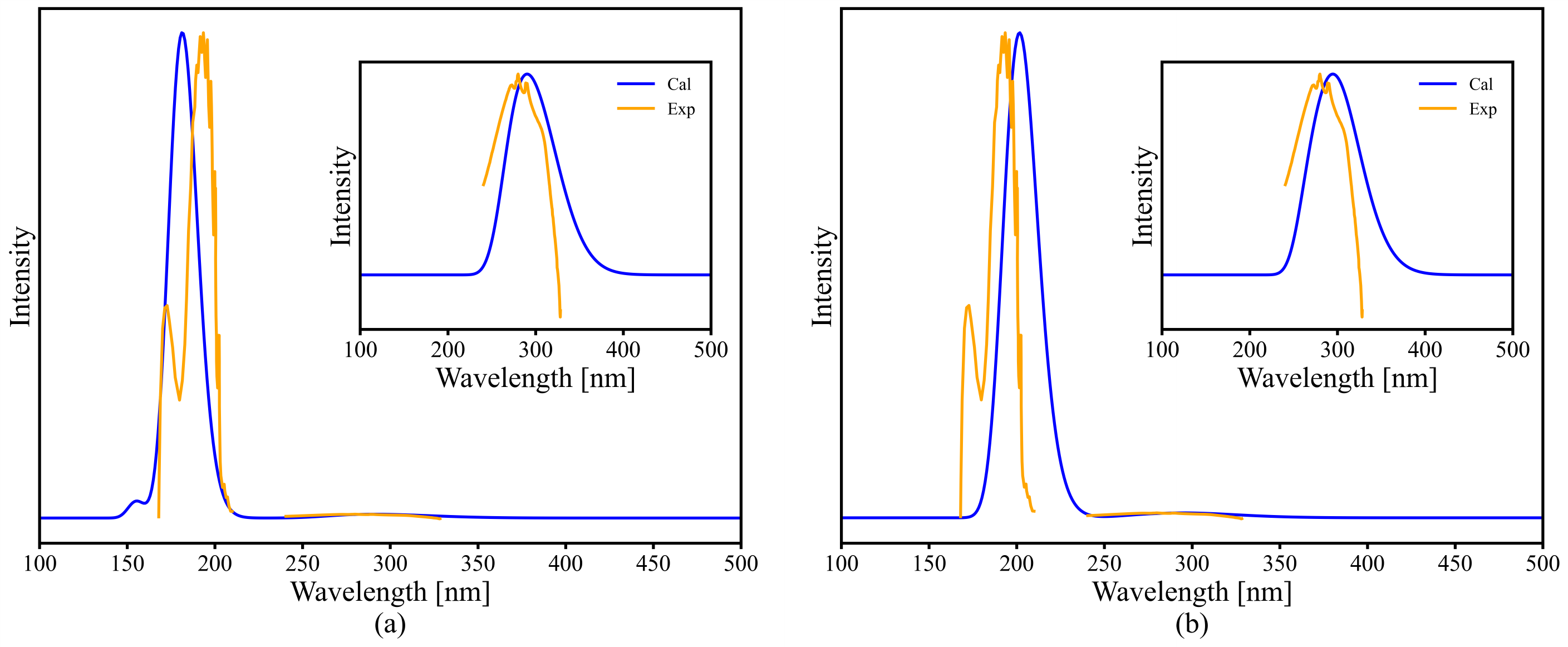}
  \caption{The absorption spectrometry based on the nuclear ensemble method in comparison with the experiments\cite{udvarhazi,hemminger1973laser}: (a) the XMS-CASPT2/def-SVPD level, (b) the XMS-CASPT2/aug-cc-pVDZ level.}
  \label{absorption}
\end{figure}

The absorption spectrum was further calculated to compare the performance of different basis sets. Fig. \ref{absorption} illustrates the simulated absorption spectrum. For the def2-SVPD basis set, the range of 390 to 230 $nm$ corresponds to a weak absorption band characterizing the $S_0 \rightarrow S_1$ transition, in which the highest peak is observed at $\sim$289 $nm$. The highest peak position exhibits $\sim$9 $nm$ shift (0.16 $eV$ deviation) in comparison with the experimental result described by the orange line in the subgraph of Fig. \ref{absorption} (a). Furthermore, the strong absorption band covers the region from 216 to 144 $nm$, and the highest peak is located at $\sim$183 $nm$. For the aug-cc-pVDZ basis set, the weak absorption band slightly shrinks to the range of 388 to 237 $nm$ with the highest peak located at $\sim$295 $nm$, while the strong absorption band covers the region from 240 to 174 $nm$, where the highest peak is located at $\sim$202 $nm$.  

Overall, the simulated absorption spectrometries can capture the main characteristics of the experimental measurement, except that the theoretical absorption band has a slight shift in the $S_0 \rightarrow S_2$ transition, in which the results of the def2-SVPD and aug-cc-pVDZ basis sets show some underestimation and overestimation, respectively. This indicates that the XMS-CASPT2 results with both the def2-SVPD and aug-cc-pVDZ basis sets are reliable to describe these low-lying excited states of cyclobutanone in the Frank-Condon region, and also confirms that the initial sampling of nuclear conditions is reasonable in the TSH dynamics, despite the significant disparity in vertical excitation energy of the $S_2$ state within two basis sets. In our preliminary dynamics explorations, it was observed that the many single-point calculations with the aug-cc-pVDZ basis set suffer from the convergence problem in the estimation of the nonadiabatic coupling vector along with the trajectory propagation. Also taking into account computational cost, the XMS-CASPT2/def2-SVPD level was utilized for subsequent nonadiabatic dynamics simulations.

\subsection{Nonadiabatic dynamics}
\begin{figure}[htbp]
  \includegraphics[width=10cm,height=8cm]{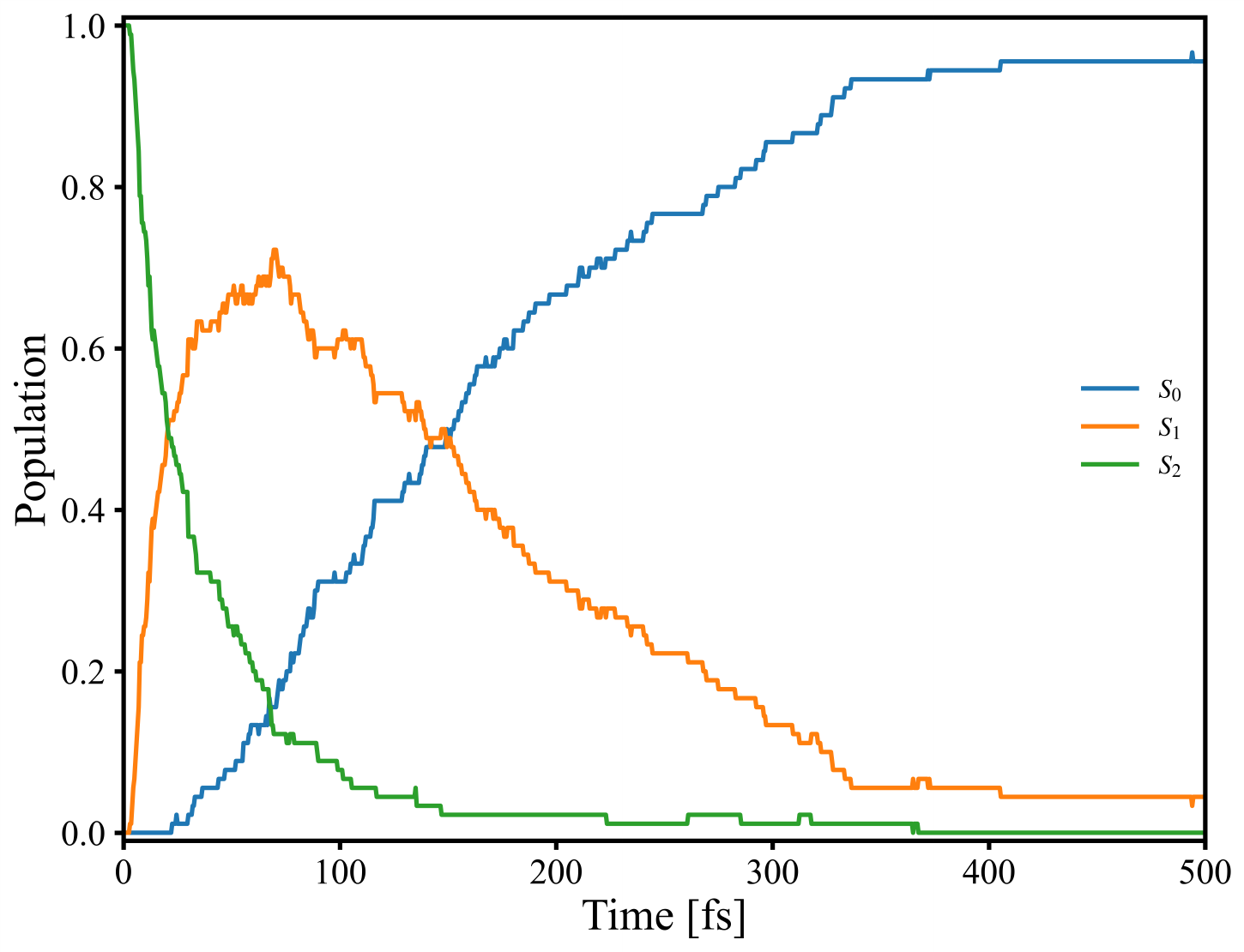}
  \caption{The time-dependent population of cyclobutanone starting from the $S_2$ state at the XMS-CASPT2/def-SVPD level.}
  \label{occ}
\end{figure}

When the dynamics starts from the $S_2$ state, the time-dependent population of cyclobutanone displays the ultrafast nonadiabatic decay dynamics, as shown in Fig. \ref{occ}. 
After excitation to the $S_2$ state, there is a rapid increase in the $S_1$ population as a result of the nonadiabatic transition from the $S_2$ to $S_1$ states. Within the first 20 $fs$, 50\% trajectories jump back to the $S_1$ state, while the $S_0$ population still remains zero at this stage. After that, the $S_0$ state begins to participate in the nonadiabatic transitions. At $\sim$90 $fs$, the $S_1$ state reaches the maximum population $\sim$72\%, and the $S_2$ population becomes similar to the $S_0$ value.
After 150 $fs$, the $S_2$ population remains consistently below 1.0\% and its contribution to the dynamics vanishes, while the population transfer between $S_1$ and $S_0$ remains. Finally, $\sim$92\% trajectories return to the $S_0$ state within 400 $fs$.

To identify which pathways are responsible for the excited state dynamics, 
Table \ref{reaction} collected the reaction ratio for various photoreaction channels. Among them, 32.6\% trajectories follow the $C2$ channel, resulting in the photodissociation products of $CH_2CO$ and $C_2H_4$. 39.1\% trajectories move toward the $C3$ channel, in which $C_3H_6$ and $CO$ were produced. In addition, 13.0\% trajectories display the hydrogen dissociation, and only 6.5\% trajectories exhibit the ring opening reaction within the time scale of the current simulation.


\begin{table}[htbp]
  \caption{The reaction products and their ratios.}
    \begin{ruledtabular}
      \begin{tabular}{ccc}
      Channel & products & Ratio \\
      \hline
      $C2$ & $CH_2CO$ + $C_2H_4$  &  32.6\%   \\
      $C3$ & $C_3H_6$ + $CO$  &  39.1\%   \\
      \textit{H dissociation} & $C_4H_5O$ + $H$  &  13.0\% \\
      \textit{ring opening} &   &  6.5\% \\
      \textit{others} &   &  8.7\%
      \label{reaction}
    \end{tabular}
  \end{ruledtabular}
\end{table}

\begin{figure}[htbp]
  \includegraphics[width=16cm,height=4.879cm]{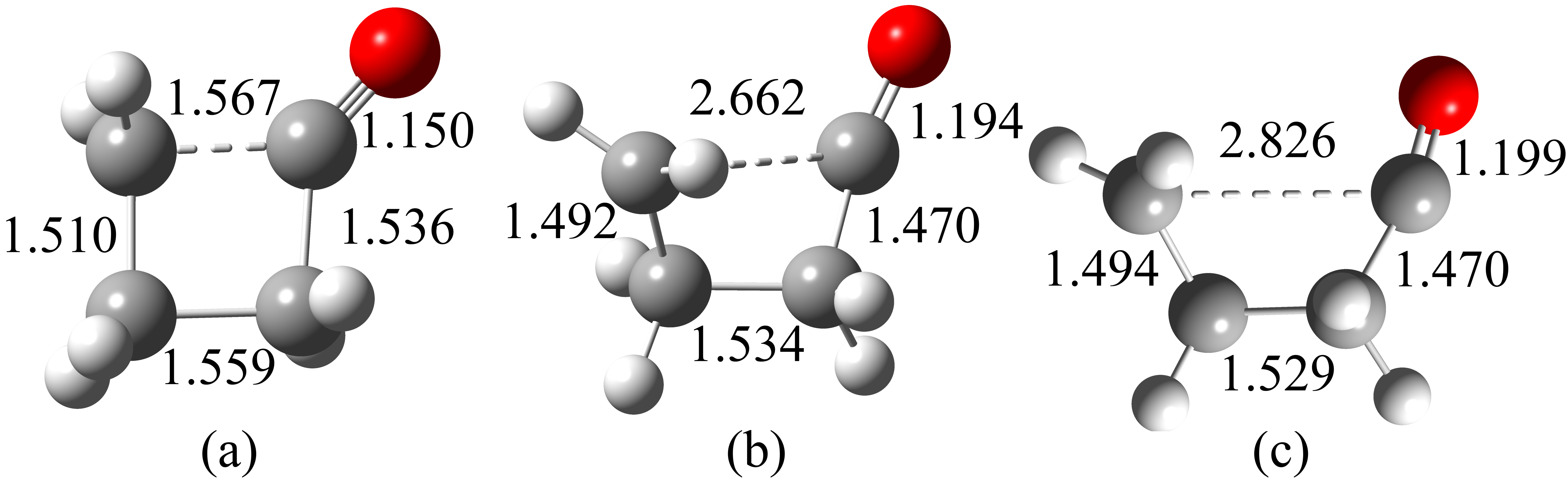}
  \caption{The optimized CI structures of cyclobutanone using the XMS-CASPT2/def-SVPD level: (a) the CI structure between the $S_2$ and $S_1$ states ($CI^{S_2/S_1}$, $E^{S_2S_1}_{CI_{S_2S_1}}-E^{S_0}_{S_0-min} = 6.980$ $eV$), (b) the CI structure between the $S_1$ and $S_0$ states ($CI_{S_1/S_0}$-1: $E^{S_1/S_0}_{CI_{S_1S_0}-1}-E^{S_0}_{S_0-min} = 3.379$ $eV$), (c) the CI structure between the $S_1$ and $S_0$ states ($CI_{S_1S_0}$-2, $E^{S_1/S_0}_{CI_{S_1S_0}-2}-E^{S_0}_{S_0-min} = 3.501$ $eV$).}
  \label{ci}
\end{figure}

In addition, the conical intersection (CI) geometries were optimized based on the hopping points observed on the trajectories, and the corresponding results are summarized in Fig. \ref{ci}. The CI structure between the $S_2$ and $S_1$ states ($CI_{S_2S_1}$) is characterized by the stretching motion of the $C(\alpha)-C$ bond with respect to $S_0$-min.
Two CI structures between the $S_1$ and $S_0$ states ($CI_{S_1S_0}-1$ and $CI_{S_1S_0}-2$) were found in the current work, which exhibit significant skeletal torsion compared to $S_0$-min. 
At both 
one of the $C(\alpha)-C$ bonds is almost broken, while the other three $C-C$ bonds remain bonded, implying that they may belong to the same CI seam.

\begin{figure}[htbp]
  \includegraphics[width=15cm,height=5.4cm]{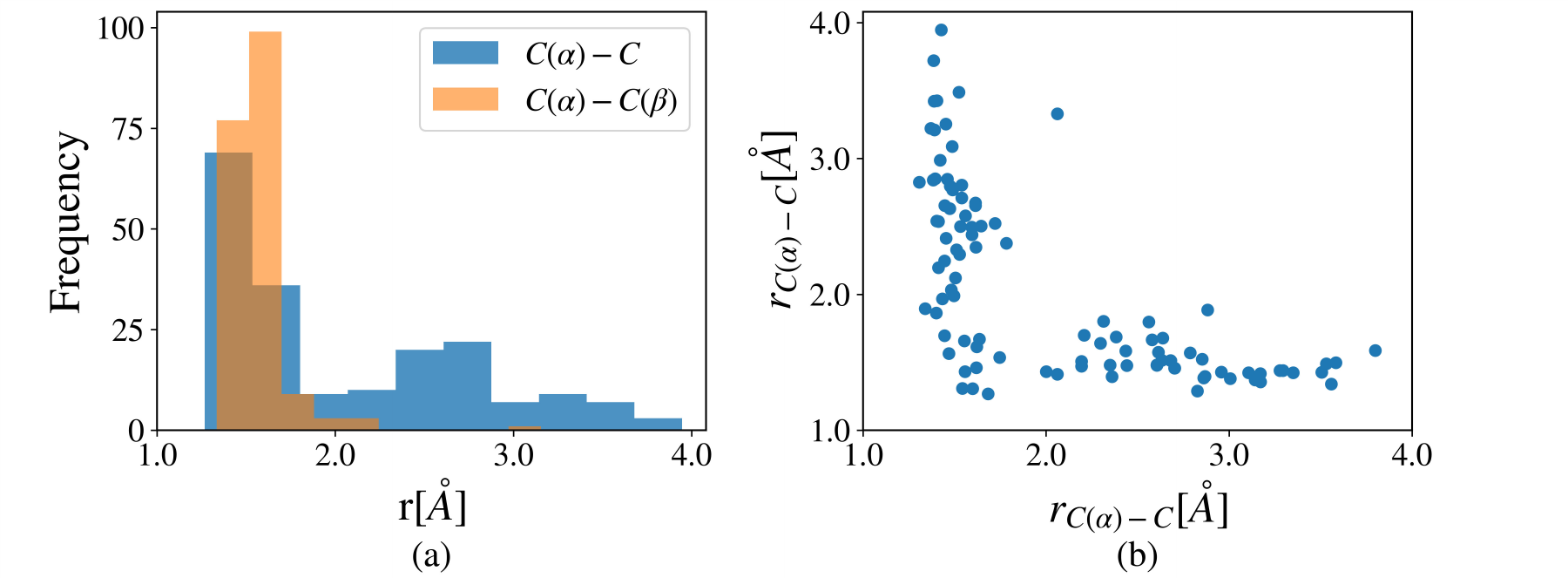}
  \caption{The characteristics of hopping structures: (a) the distributions of the $C(\alpha)-C$ and $C(\alpha)-C(\beta)$ distances at hopping points for the $S_1\rightarrow S_1$ transition, (b) the correlation of different $C(\alpha)-C$ bonds.}
  \label{distribution}
\end{figure}

The distributions of the $C(\alpha)-C$ and $C(\alpha)-C(\beta)$ distances at the $S_1\rightarrow S_0$ hops are given in Fig. \ref{distribution} (a) and (b).
All $C(\alpha)-C(\beta)$ distances remain in the short-distance domain (Fig. \ref{distribution} (a)), indicating that both two $C(\alpha)-C(\beta)$ bonds remain.
As contrast, the $C(\alpha)-C$ bond shows two peaks: a narrow peak is found at a short distance 
and a board peak appears in the long distance domain.
Combined with the distribution of two different $C(\alpha)-C$ bonds (Fig. \ref{distribution} (b)), it is clear that only one of the $C(\alpha)-C$ bonds breaks before the internal conversion to the ground state. 
Overall, the hopping distribution is consistent with two $CI_{S_1S_0}$ structures.  





\subsection{GUED  spectrum}
\begin{figure}[htbp]
  \includegraphics[width=16cm,height=7.4cm]{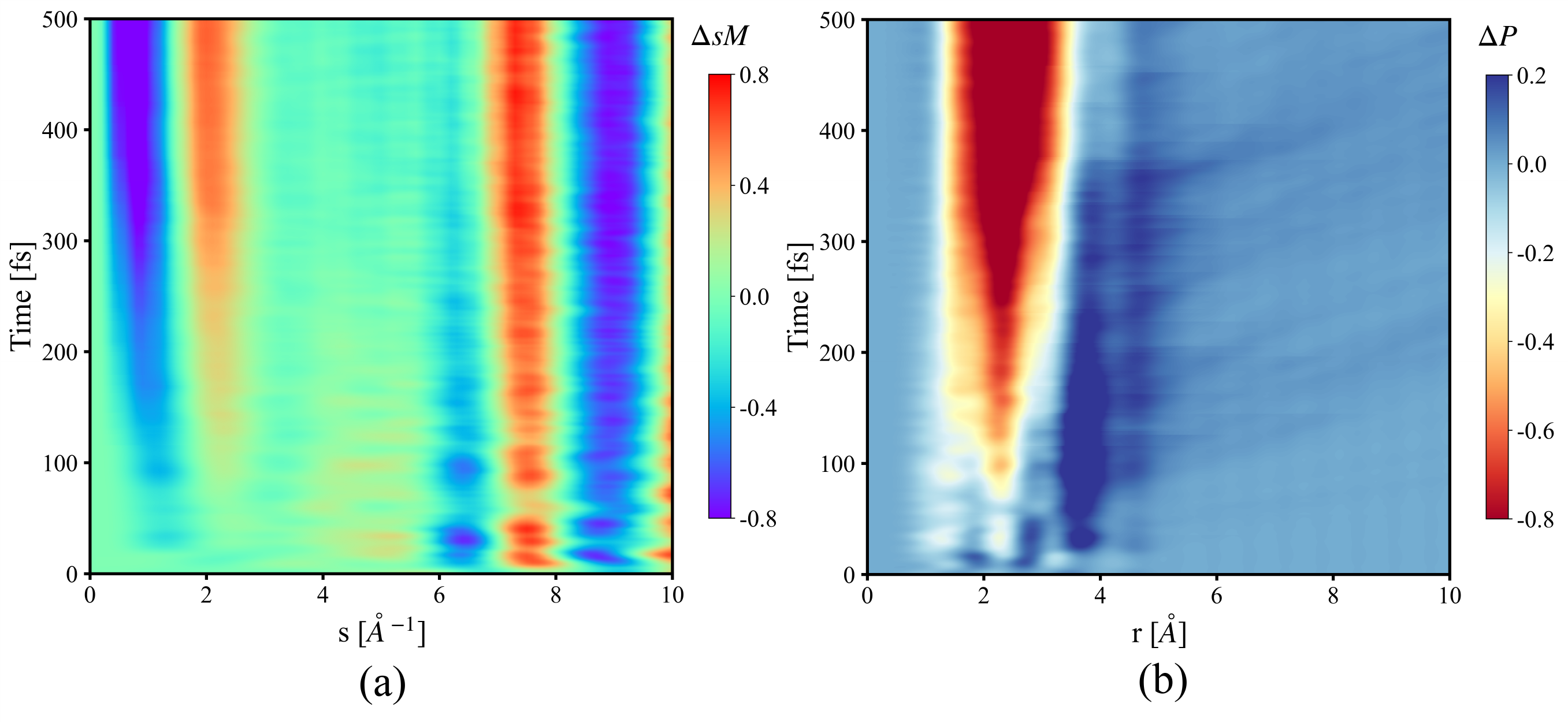}
  \caption{The time-dependent diffraction signals of cyclobutanone: (a) $\Delta sM(s, t)$ in terms of the $S_0$-min as reference, (b) $\Delta P(r, t)$ in terms of the $S_0$-min as reference.}
  \label{diffraction}
\end{figure}

The scattering signals were stimulated using the IAM framework on the basis of TSH trajectories. Fig. \ref{diffraction} shows the averaged time-dependent diffraction signals of cyclobutanone, including the differences of the modified signal $\Delta sM(s, t)$ and the distribution function $\Delta P(r, t)$. In $\Delta sM(s, t)$, all signals appear after the first several femtoseconds and persist throughout the dynamics. The negative contributions exist in the $0.3$ $\overset{\circ}{A}$$^{-1}< s < 1.5$ $ \overset{\circ}{A}$$^{-1}$, $ 5.8$ $\overset{\circ}{A}$$^{-1} < s < 6.8$ $\overset{\circ}{A}$$^{-1}$ and $8.2$ $\overset{\circ}{A}$$^{-1} < s < 9.6 $ $\overset{\circ}{A}$$^{-1}$ regions, in which the differences of scattering intensities continue to increase over time in the first and third regions, while decay over time in the second region. The positive contributions mainly occur in the $1.6$ $\overset{\circ}{A}$$^{-1} < s < 2.8$ $\overset{\circ}{A}$$^{-1}$ and $6.8$ $\overset{\circ}{A}$$^{-1} < s < 8.2$ $\overset{\circ}{A}$$^{-1}$ areas, where the difference of the scattering intensity increases with time in the former.
In $\Delta P(r, t)$, the difference in the pair distribution function is also very weak for the first several femtoseconds, which corresponds to the $S_2 \rightarrow S_1$ relaxation in population dynamics. Then, the negative contributions were mainly observed in the $1.0$ $\overset{\circ}{A} < r < 3.0$ $\overset{\circ}{A}$ region, where the signal difference increases with time. Simultaneously, the positive contributions occur at $r > 3.5$ $\overset{\circ}{A}$, and the main peak occurs at $r \approx 3.6$ $\overset{\circ}{A}$, which remains essentially unchanged unitl 380 $fs$, and then slightly decays in the remaining time. In addition, a relatively weaker positive band occurs at $r \approx 4.5$ $\overset{\circ}{A}$, and the signal difference also becomes larger with time at the longer distance. The positive signals were also found in the long-distance domain, which are characterized by low intensities and long tails. Overall, 
the difference in the pair distribution function clearly demonstrates that some chemical bonds vanish in the short distance region ($1.0$ $\overset{\circ}{A} < r < 3.0$ $\overset{\circ}{A}$), and more atomic pairs are formed in the region of $3.5$ $\overset{\circ}{A} < r < 4.0$ $\overset{\circ}{A}$ with time being. Moreover, the dissociation dynamics takes place, and some atomic pairs show very large distances.     


\begin{figure}[htbp]
  \includegraphics[width=16cm,height=7.25cm]{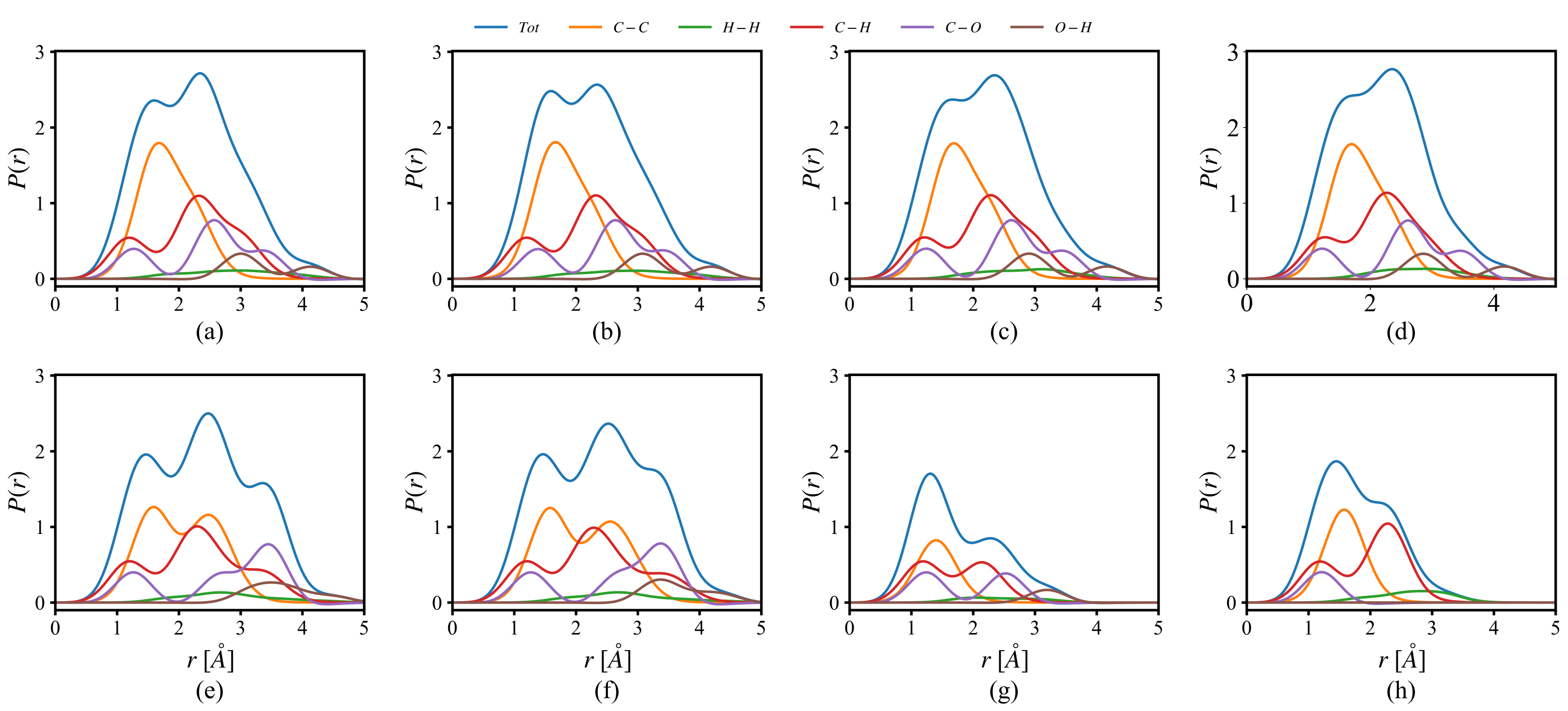}
  \caption{The diffraction signals of cyclobutanone at different structures: (a) $S_0$-min, (b) $S_1$-min, (c) $S_2$-min, (d) $CI_{S_2S_1}$, (e) $CI_{S_1S_0}-1$, (f) $CI_{S_1S_0}-2$, (g) $C2$ products and (h) $C3$ products.}
  \label{single}
\end{figure}

To understand different signals and establish the connection with geometric movements, some important structures, including $S_0$-min, $S_1$-min, $S_2$-min, $CI_{S_2S_1}$, $CI_{S_1S_0}$-1 and $CI_{S_1S_0}$-2, were selected to calculate the real-space distribution function $P(r)$, and the corresponding results are collected in Fig. \ref{single}. For the overall $P(r)$ result, $S_0$-min, $S_2$-min and $CI_{S_2S_1}$ exhibit similar patterns. Taking the $S_0$-min signal as an example, two main peaks appear in the total $P(r)$ signal, one is located at $\sim$1.5 $\overset{\circ}{A}$, which is assigned as the $C-C$ contribution compared to other components shown in Fig. \ref{single}(a). The other peak occurs at $\sim$2.5 $\overset{\circ}{A}$, in which the $C-C$, $C-H$ and $O-H$ contributions take dominant roles here. 
For the individual $P(r)$ result, the $C-C$ contribution only has a single board peak with the center position locating $\sim$1.5 $\overset{\circ}{A}$ at the current resolution, 
which is in principle consistent with the $S_0$-min distance between neighboring $C$ and $C$ atoms.
The contribution between two non-bound $C-C$ atomic pairs is not obviously observed in the current very broad GUED signal, and we expect that they can be distinguished by a further adjustment of the resolution. The $C-H$ contribution mainly has two peaks, one occurs at $\sim$1.2 $\overset{\circ}{A}$, which corresponds to the length of the $C-H$ bond; the other appears at $\sim$2.5 $\overset{\circ}{A}$, which is related to the distance between $C$ and $H$ connecting to the other $C$ atom. The $O-H$ contribution mainly shows two peaks, one is located at $\sim$3.0 $\overset{\circ}{A}$, which corresponds to the distance between $O$ and $H$ connecting to the $C(\alpha)$ atom. The other occurs at $\sim$4.2 $\overset{\circ}{A}$, which is related to the distance between $O$ and $H$ that connects to the $C(\beta)$ atom. The $C-O$ contribution exhibits three peaks corresponding to the $C-O$, $C(\alpha)-O$ and $C(\beta)-O$ pairs, respectively. All peak intensities of each component are consistent with the number of the corresponding atomic pair. 

Compared to the $S_0$-min signal, the overall peak profile of $P(r)$ at $S_1$-min is similar, while the peak intensities at $\sim$1.5 and $\sim$2.5 $\overset{\circ}{A}$ show a slight increase and decrease, respectively, resulting in comparable heights of these two peaks.
The $P(r)$ signal at $CI_{S_2S_1}$ also exhibits an analogous pattern, which is consistent with its slight geometric distortion with respect to $S_0$-min. 
Differently, the $P(r)$ signals at both $CI_{S_1S_0}$-1 and $CI_{S_1S_0}$-2 display three peaks, 
and the $C-C$ peak at short distance ($<2$ $\overset{\circ}{A}$) becomes weaker. A new peak appears at  $\sim$3.5 $\overset{\circ}{A}$, which is mainly caused by the $C-O$ contribution. 
The above features indicate that some $C-C$ bonds break and some $C-O$ atomic pairs appear at the longer distance. 
For the final $C2$ and $C3$ products at the dissociation limits, the overall peak positions do not change in this case, while the total $P(r)$ intensities show obviously decay and the peak signals at both $\sim$1.5 and $\sim$2.5 $\overset{\circ}{A}$ reduce significantly.
In particular, the intensity of the second peak becomes almost one third of its original value for the $C_2$ product.
For the $C_3$ channel, both peaks also become weaker, while the second peak at $\sim$2.5 $\overset{\circ}{A}$ decays to almost one half of its value compared to that at $S_0$-min.
If we compare the peak profiles for two dissociation channels, their different intensities of the peak at $\sim$2.5 $\overset{\circ}{A}$   reflect different bond statues of their dissociation products. 

%

Therefore, after combining the above $\Delta sM(r, t)$ and $\Delta P(r, t)$, the increasing signal at above 3 $\overset{\circ}{A}$ and the weakening signal intensities at 2.05 and 1.89 $\overset{\circ}{A}$ can be ascribed to the ring opening process and subsequent dissociation into different molecular products.

\begin{figure}[htbp]
  \includegraphics[width=16cm,height=13.45cm]{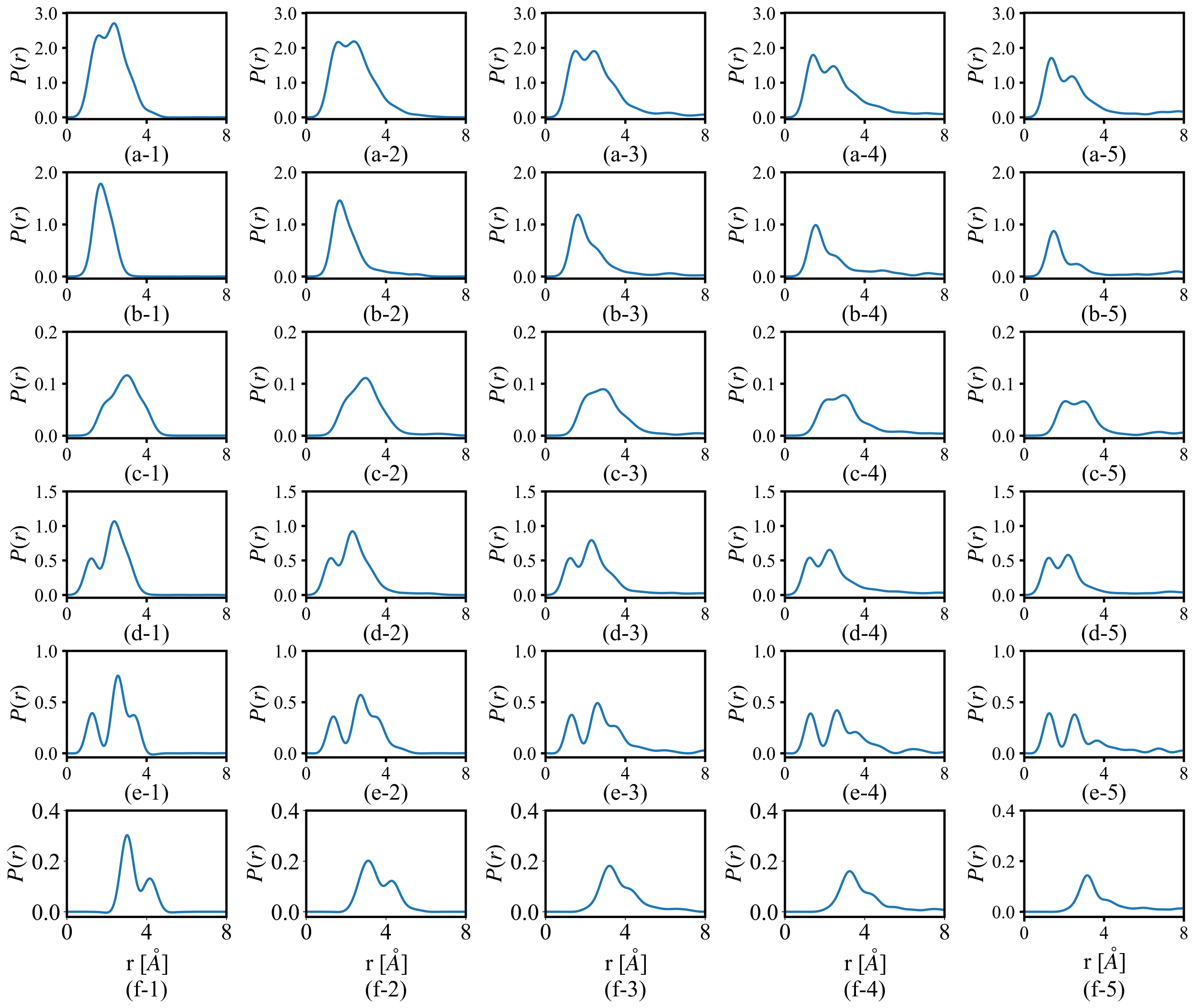}
  \caption{The time-dependent diffraction signals of cyclobutanone for $C2$ channel: (a) total signal, (b) $C-C$ contribution, (c) $H-H$ contribution, (d) $C-H$ contribution, (e) $C-O$ contribution, (f) $O-H$ contribution, the indices 1-5 correspond to 0, 100, 200, 300, 400 $fs$, respectively.}
  \label{time_c2}
\end{figure}

\begin{figure}[htbp]
  \includegraphics[width=16cm,height=13.45cm]{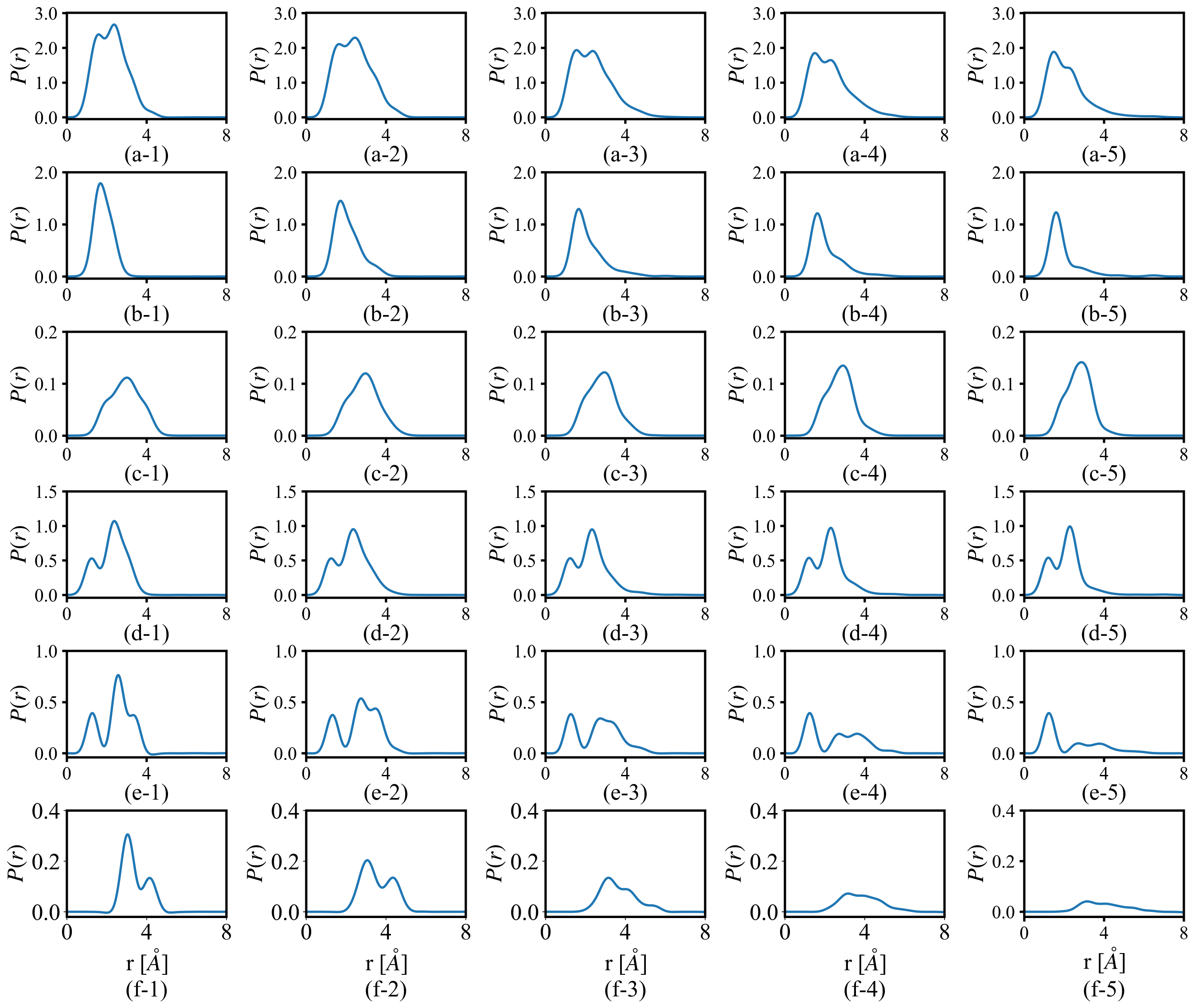}
  \caption{The time-dependent diffraction signals of cyclobutanone for $C3$ channel: (a) total signal, (b) $C-C$ contribution, (c) $H-H$ contribution, (d) $C-H$ contribution, (e) $C-O$ contribution, (f) $O-H$ contribution, the indices 1-5 correspond to 0, 100, 200, 300, 400 $fs$, respectively.}
  \label{time_c3}
\end{figure}

\begin{figure}[htbp]
  \includegraphics[width=16cm,height=13.45cm]{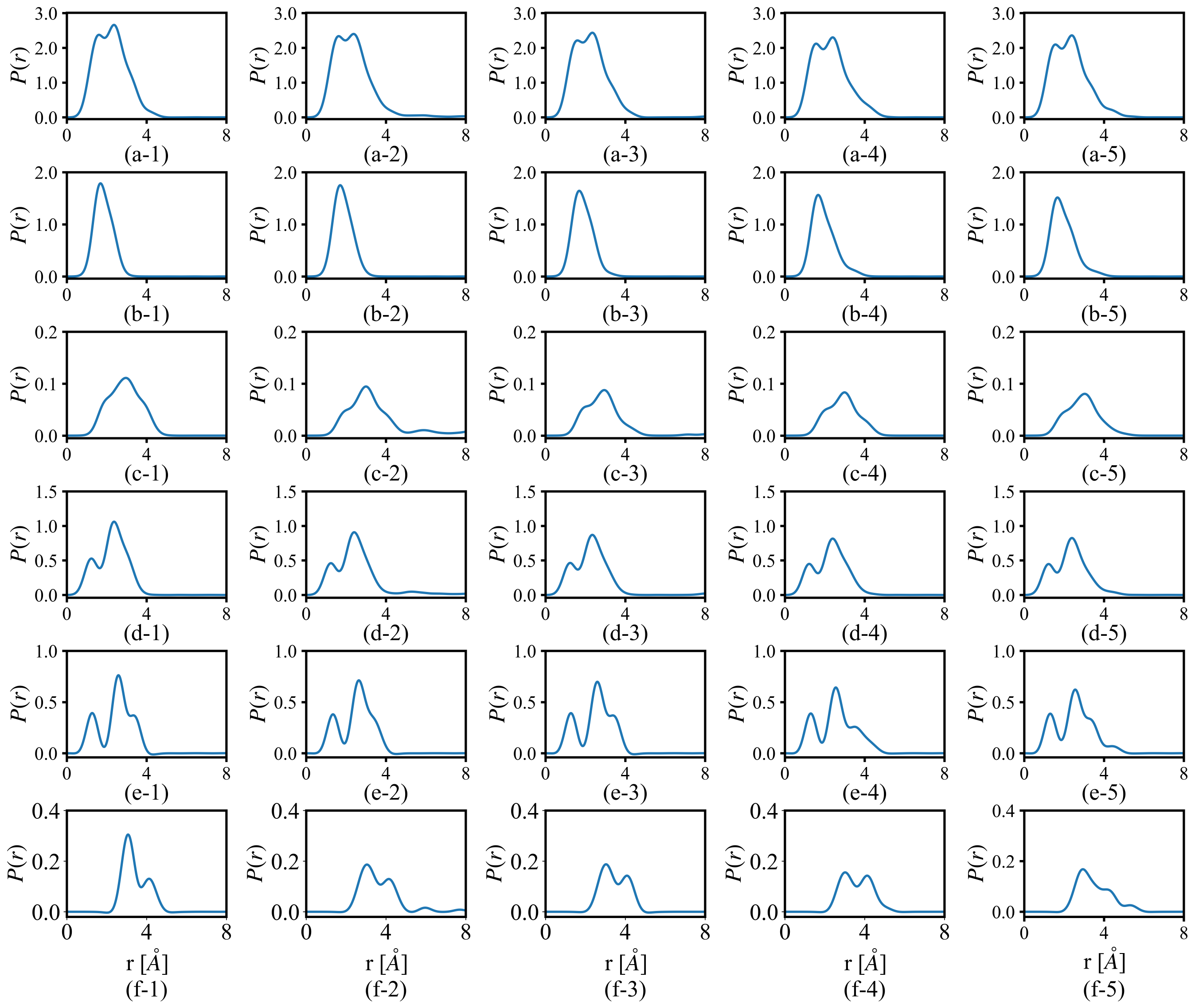}
  \caption{The time-dependent diffraction signals of cyclobutanone for $H$ dissociation channel: (a) total signal, (b) $C-C$ contribution, (c) $H-H$ contribution, (d) $C-H$ contribution, (e) $C-O$ contribution, (f) $O-H$ contribution, the indices 1-5 correspond to 0, 100, 200, 300, 400 $fs$, respectively.}
  \label{time_h}
\end{figure}

Based on the analysis presented above, the diffraction signals for different reaction channels with ratios greater than 10\% were obtained by averaging relevant trajectories to investigate the time-dependent characteristics, respectively. The corresponding results are summarized in Figs. \ref{time_c2}, \ref{time_c3} and \ref{time_h}. 

For the $C2$ channel, 
there are two peaks in the total GUED spectra at time zero, the long-distance peak is slightly higher than the short-distance one, which is consistent with the molecular geometry at the Franck-Codon region (Fig. \ref{single}(a)). Within 100 fs, both peaks in total $P(r)$ exhibit visible decays, and their intensities become similar. 
If we compare Fig. \ref{single} and Fig. \ref{time_c2}, it is clear that the system decays to the $S_1$ state within 100 fs, and this time scale is consistent with the population dynamics in Fig. \ref{occ}. Afterward, both peaks in the total GUED signals decay, particularly for the long-distance one. In addition, a long tail is observed in the signal, indicating that the dissociation takes place.
More details are obtained by the analyses of all  
individual $P(r)$. 
For the $C-C$ distance contribution (Fig. \ref{time_c2}(b-1 to b-5)), the strong decrease of the peak intensity and the appearance of the visible tail up to the very long-distance region indicate that some $C-C$ bonds break. 
In the $C-O$ distance distribution, three peaks are given at time zero due to the existence of three types of $C-O$ atomic pairs ($C-O$, $C(\alpha)-O$ and $C(\beta)-O$), and the intensity of the middle peak ($C(\alpha)-O$ pair) is almost twice that of two other peaks due to the existence of two $C(\alpha)$ atoms. With time being, it is clear that the intensity of the middle peak decreases, indicating that one of the $C(\alpha)-O$ atomic pairs tends to vanish and their distance becomes longer. The peaks at other locations do not change, so the bond-breaking process is asynchronous. 
%
%
After 300 fs, two peaks corresponding to the $C-O$ and $C(\alpha)-O$ pairs become similar, indicating that the ratio of the $C$ and $C(\alpha)$ atoms is almost 1:1 in the vicinity of the $O$ atom. 
Compared to the initial signals, it is clear that only one $C(\alpha)-O$ atomic pair remains. In addition, the $C(\beta)-O$ atomic pair tends to vanish at the dissociation limit.
Taking into account the long tail in Figs. \ref{time_c2} (e-4) and (e-5) further, it is clear that the final product should contain the $C-C-O$ backbone.
As each $C(\alpha)$ or $C(\beta)$ atom connects to two $H$ atoms at time zero, we see two peaks in the $O-H$ distance distribution, in which
the short-distance $O-H(-C(\alpha))$ pair peak is much higher than the long-distance $O-H(-C(\beta))$ one.
However, the long-distance peak vanishes after 200 fs, indicating the vanishing of the $O-H(-C(\beta))$ pair. The short-distance peaks become much lower than the original one, implying that only the reduction of the $O-H(-C(\alpha))$ pairs. Overall, the above observations are highly consistent with the whole dynamics processes, {i.e.} the system first decays via $CI_{S_1S_0}$ to the ground state and afterward the dissociation takes place to give the $CH_2CO$ molecule.


For the $C3$ channel, the total GUED signals in the early time show similar features to those of the $C2$ channel, while the pattern is different.
Only the minor change of the $C-C$ distribution peak with time is observed, and its intensity becomes slightly weaker. 
The distribution of the $C-H$ distance also shows a very small dependence on time, indicating that all $C-H$ pairs remain almost unchanged. This is consistent with the fact that the $C_3H_6$ backbone does not break. 
For the $C-O$ distance distribution, it is clear that the long-distance (> 2 $\overset{\circ}{A}$) distribution becomes much weaker and the short-distance distribution part remains unchanged. In other word, the long-distance $C-O$ pairs ($C(\alpha)-O$ and $C(\beta)-O$) become less, while the $C-O$ bond keeps. This indicates that the cleavage takes place for some $C-C$ bonds.  
At 400 fs, we noticed that only the $C-O$ pair with the shortest distance exists. The weak long tail indicates that other $C-O$ pairs almost vanish and their distances become very long. For the $O-H$ distance distribution, initially two peaks are observed, consistent with the signals at $S_0$-min. With time being, almost all peaks disappear, indicating that all $O-H$ pairs vanish. 
%
Although the GUED spectra at 400 fs do not yet match those in the dissociation limit, it is clear that the $CO$ and $C_3H_6$ compounds are formed in this channel.

%

For the $H$ dissociation channel, the total $P(r)$ signal exhibits a slight change, as shown in Fig. \ref{time_h}, and the major difference is reflected by the individual signals relevant to the $H-H$, $C-H$ and $O-H$ contributions. This indicates that the whole backbone remains similar, while all atomic pairs involving the $H$ atom tend to become smaller. This partially reflects that the $H$ dissociation takes place.
  



\newpage
\section{CONCLUSIONS}
In this work, the ultrafast electron diffraction image of cyclobutaone was predicted to perform a double-blind test of accuracy in excited-state simulations, with respect to the benchmark data obtained by upcoming MeV-GUED experiments at the Stanfold Linear Accelerator Laboratory. For this purpose, the \textit{ab initio} nonadiabatic dynamics was simulated by the trajectory surface hopping method at the XMS-CASPT2 level starting from the $S_2$ state, and the time-resolved GUED signals were calculated directly within the IAM framework.

The simulation results show that the ultrafast nonadiabatic dynamics occurs in cyclobutaone, and two $C2$ and $C3$ channels take dominant roles in the dynamics trajectories.
The time-dependent bond evolution of these channels can be well characterized by the GUED signals. For example, we can clearly observe the $CC$ bond cleavage and the $CH_2CO$ formation in the $C_2$ channel.
For the $C-3$ channel, the formation of the $CO$ and $C_3H_6$ compounds is clearly identified. In this sense, the GUED spectra can help to clarify the key atomic motions in the photoinduced dynamics of cyclobutaone. 
%
With the development of the time-resolved GUED experimental facilities and the novel computational theories, we believe that the combination of both sides will provide a unique tool for understanding the nonadiabatic dynamics of realistic molecules at the all-atomic level with both high temporal and fine spatial resolutions.


\begin{acknowledgments}
This work was supported by NSFC projects (Nos. 22333003, 22361132528, 21933011), and the Opening Project of Key Laboratory of Optoelectronic Chemical Materials and Devices of Ministry of Education, Jianghan University (JDGD-202216). The authors thank the Supercomputing Center, Computer Network Information Center, Chinese Academy of Sciences; National Supercomputing Center in Shenzhen for providing computational resources.
\end{acknowledgments}

\section*{Data Availability Statement}
The data supporting this study are available from the corresponding author upon reasonable request.


\nocite{*}
\bibliography{aipsamp}

\end{document}